\documentclass[preprint,showpacs,preprintnumbers,amsmath,amssymb,nofootinbib]{revtex4}
\usepackage{graphicx}
\def \be {\begin{equation}}

\def \ee {\end{equation}}
\def \bea {\begin{eqnarray}}
\def \eea {\end{eqnarray}}

\usepackage{graphicx}
\usepackage{dcolumn}
\usepackage{bm}

\begin{document}

\title{Simplified Chaplygin Gas: Deriving $H_0$ From Ages of Old High Redshift Objects and Baryon Acoustic Oscillations}

\author{R. C. Santos} \email{cliviars@astro.iag.usp.br}
\author{J. F. Jesus} \email{jfernando@astro.iag.usp.br}
\vskip 0.5cm \affiliation{Departamento de Astronomia, Instituto de Astronomia, Geof\'{i}sica e Ci\^{e}ncias Atmosf\'ericas, Universidade de S\~ao Paulo, 05508-900 S\~ao Paulo, SP, Brasil}

\pacs{Age of the Universe; BAO; Chaplygin Gas.}
\begin{abstract}
The discovery that the expansion of the Universe is accelerating is the most challenging problem of modern cosmology. In the context of general relativity, there are many dark energy candidates to explain the observed acceleration.
In this work we focus our attention on two kinds of simplified Chaplygin gas cosmological accelerating  models recently  proposed in the
literature. In the first scenario, the simplified Chaplygin gas works
like a Quintessence model while in the second one, it plays the role of a Quartessence 
(an unification of the dark sector). Firstly, in order to limit the free parameters of both models, we  discuss the age of high
redshift objects with special emphasis to the old quasar APM
08279+5255 at $z = 3.91$. The basic finding is that this old high redshift object constrain severely the simplified Chaplygin cosmologies. 
Secondly, through a joint analysis involving the baryon acoustic oscillations (BAO) and
a sample of old high redshift galaxies (OHRGs) we also estimate the value
of the Hubble parameter, $H_0$. Our approach suggests 
that the combination of these two independent phenomena provides an
interesting method to constrain the Hubble constant.

\end{abstract}

\maketitle

\section{Introduction}
\hspace{0.5cm}

An impressive amount of different astrophysical data are suggesting 
that the observed Universe can be represented  by an accelerating  spatially flat cosmology driven by an exotic component 
sometimes called dark energy \cite{Riess,Sperg07}. In addition to the cosmological constant,  
the most interesting candidates for describing this dark energy component are: a vacuum
decaying energy density, or a time varying $\Lambda(t)$ \cite{OzTa87}, 
the so-called ``X-matter" \cite{turner97}, a relic
scalar field rolling down its potential \cite{PR03}, and a Chaplygin Gas \cite{kamen}. 
Some recent review articles discussing the history, interpretations, as
well as the major difficulties of such candidates have also been
published in the last few years \cite{review}.

On the other hand, some studies have pointed out that the age of old high redshift objects (at moderate and 
high redshifts) is a powerful physical constraint to cosmological models \cite{old,Krauss97,HasingerEtAl02,LA,Age}. 
For old high redshift galaxies (OHRGs), for instance, their ages  can be inferred by measuring the magnitude in different
bands and then using stellar evolutionary codes to choose the model
that reproduces the observed colors \cite{old,Krauss97}. 
In principle, for old high redshift objects, the derived ages can put tight constraints on the 
physical parameters or even rule out many kinds of dark energy models.

In this article, we discuss some  observational constraints on two different classes of Chaplygin type accelerating 
cosmologies coming  from  the existence of the  APM
08279+5255, an old quasar at high redshift located  at $z = 3.91$ which 
has an estimated age of 2-3 Gyr \cite{HasingerEtAl02}. As reported by  many authors \cite{LA},  
the existence of this object is not compatible with many dark energy models 
unless the present accepted values  of $\Omega_M$ and ${H}_0$ be further revised. We also derive new constraints on the Hubble constant
$H_0$, by using a joint analysis involving different OHRGs
samples and the current SDSS measurements of the baryon acoustic
peak \cite{Eisenst05}. The BAO
signature comes out because the cosmological perturbations excite
sound waves in the relativistic plasma, thereby producing the
acoustic peaks in the early universe. Actually, Eisenstein {\it et al.} \cite{Eisenst05} presented
the large scale correlation function from the Sloan Digital Sky
Survey (SDSS) showing clear evidence for the baryon acoustic peak at
$100h^{-1}$ Mpc scale, which is in excellent agreement with the WMAP
prediction from CMB data. We stress that the
Baryon Acoustic Oscillations (BAO) method is independent of the
Hubble constant $H_0$ although being heavily dependent on the value of
$\Omega_M$. In this way, 
it contributes indirectly to fix the
value of $H_0$ since it breaks
the degeneracy on the mass density parameter, $\Omega_M$ \cite{CLM07}. 
As we shall see, both set of data
(OHRGs sample and BAO signature) provide an interesting method to
constrain the Hubble constant.

\section{THE MODELS}

\subsection{Simplified Chaplygin Quintessence Cosmology}

The basic idea underlying  the simplified Chaplygin gas concept (from now on
SC-gas) either the Quintessence \cite{scg} or Quartessence \cite{scg1} version,  is to reduce the number of free parameters of the generalized C-gas \cite{kamen,GCG}, however, maintaining the physically interesting properties. In the Quintessence case it behaves as a pressureless fluid (nonrelativistic matter) at
high-$z$ while, at late times, it approaches the quintessence
behavior. The generalized C-gas has as equation of state (EOS):
\begin{equation}
 p_C=-A/\rho_C^\alpha,
\end{equation}
which leads to  the following adiabatic sound speed:
\begin{equation}
 v_s^2=\frac{dp}{d\rho}=\alpha A/\rho_C^{1+\alpha}.
\end{equation}

By using the definition $A_s\equiv A/\rho_{C0}^\alpha$, the present C-gas adiabatic sound speed reads $v_{s0}^2=\alpha A_s$. Thus,  if we want to require the positiviness of $v_{s0}^2$ and also want to reduce the number of free parameters of the C-gas, the simplest way of doing that is imposing $A_s=\alpha$. With this we have that $v_{s0}^2=\alpha^2$, thus assuring a natural positiviness for $v_{s0}^2$. Furthermore, in order to  guarantee that $v_s\le c$, the $\alpha$ parameter  it will be restricted on the  interval $0<\alpha\le1$ (for more details see Lima, Cunha and Alcaniz \cite{scg,scg1}). As a consequence, the simplified Quintessence Chaplygin gas is fully characterized only by two parameters: $\alpha$ and $\Omega_C$.  As one may check, for a spatially flat Friedmann-Robertson-Walker geometry the dimensionless Hubble parameter is given by 
\begin{equation}
\label{EzQuint}
E(z) = \left[\Omega_M (1+z)^3 + \Omega_C \left[\alpha + (1 - \alpha)
(1+z)^{3(\alpha + 1)}\right]^{\frac{1}{\alpha + 1}}\right]^{1/2},
\end{equation}
where $E(z) \equiv H(z)/H_0$, and $\Omega_C=1-\Omega_M$. In particular, this means that the$\alpha$  parameter  is actually the unique unknown constant related to this SC-gas model. The free parameters of such cosmologies are reduced to  $\Omega_M$, and the present rate of expansion, $H_0$.

\subsection{Simplified Quartessence Cosmology}

In the Quartessence version of the SC-gas, the model maintain all the physical requirements of the Quintessence scenario, with the additional feature of unifying dark sector (dark matter and dark energy). In such a scenario, besides the SC-gas the unique nonvanishing contribution (radiation is important only at early times) comes from baryonic component which is tightly constrained by big-bang nucleosynthesis (BBN)  and the temperature anisotropies of the cosmic background radiation (CMB).

In this case, it is readily seen that the dimensionaless Hubble parameters is given by
\begin{equation}
\label{EzQ4}
E(z) = \left[\Omega_b (1+z)^3 + \Omega_{Q4} \left[\alpha + (1 - \alpha)
(1+z)^{3(\alpha + 1)}\right]^{\frac{1}{\alpha + 1}}\right]^{1/2},
\end{equation}
where $\Omega_b$ is the baryon contribution, $\Omega_{Q4}$ stands for the SC-gas as Quartessence contribution. Note that $\Omega_{Q4}=1-\Omega_b$ because we have assumed  spatial flatness. Therefore, since $\Omega_b$ is quite well determined, one may conclude that only two parameters  remain to be constrained in this
unified dark matter/energy scenario, namely, $\alpha$ and $H_0$.

\section{OBSERVATIONAL CONSTRAINTS}
\subsection{Age-Redshift Test}
Let us now discuss the age-redshift test in the above discussed backgrounds. For both cases, the expression for the age relation $t_z$ takes the following form:
\begin{equation}
\label{tz}
H_0 t_z = \int^{z}_{0}\frac{dz'}{(1 +z')E(z')},
\end{equation}
where $E(z)$ is given by (\ref{EzQuint}) or (\ref{EzQ4}) for the Quintessence and Quartessence versions, respectively. Note that for $\Omega_M = 1$
the above expression reduces to the well known result for
Einstein-de Sitter model (CDM, $\Omega_M = 1$) for which $t_z =
\frac{2}{3} H^{-1}_{0} (1 + z)^{-3/2}$. As one may conclude from the
above equation, limits on the cosmological parameters $\Omega_M$ (or $\Omega_b$), $\alpha$ and
$H_0$ (or equivalently $h\equiv H_0({\rm kms^{-1}Mpc^{-1}})/100$), can be derived by fixing $t_z$ from
observations. Note also that the age
parameter, $T_z = H_0 t_z$, depends only on the product of the two
quantities $H_0$ and $t_z$, which are usually estimated from
completely independent methods \cite{Peebles}. The age-redshift test is given by the condition
\begin{equation}
\frac{T_z}{T_q} = \frac{f(\Omega_{M,b}, \alpha, z)}{H_0 t_q} \geq
1,
\end{equation}
where $t_q$ is the age of an arbitrary object, say, a quasar or a
galaxy at a given redshift $z$ and $f(\Omega_{M,b},\alpha, z)$ is
the dimensionless factor given by the $rhs$ of Eq. (\ref{tz}). For each object, the
denominator of the above equation defines a dimensionless age
parameter $T_q = H_0 t_q$. In particular, for the 2.0-Gyr-old quasar
at $z = 3.91$ estimated by Hasinger et al. (2002) \cite{HasingerEtAl02} yields $T_q = 2.0H_0$Gyr. In addition,  for the most recent
determinations of  the Hubble parameter, $H_0 = 72 \pm 8$ ${\rm{km
s^{-1} Mpc^{-1}}}$ as given by the Hubble Space Telescope Key Project \cite{Freedman01}, the age parameter take values on the
interval $0.131 \leq T_q \leq 0.163$. It follows that $T_q \geq
0.131$. Therefore, for a given value of $H_0$, only models having an
expanding age bigger than this value at $z = 3.91$ will be
compatible with the existence of this object.

\begin{figure}[t]
\centering
\includegraphics[width=82mm]{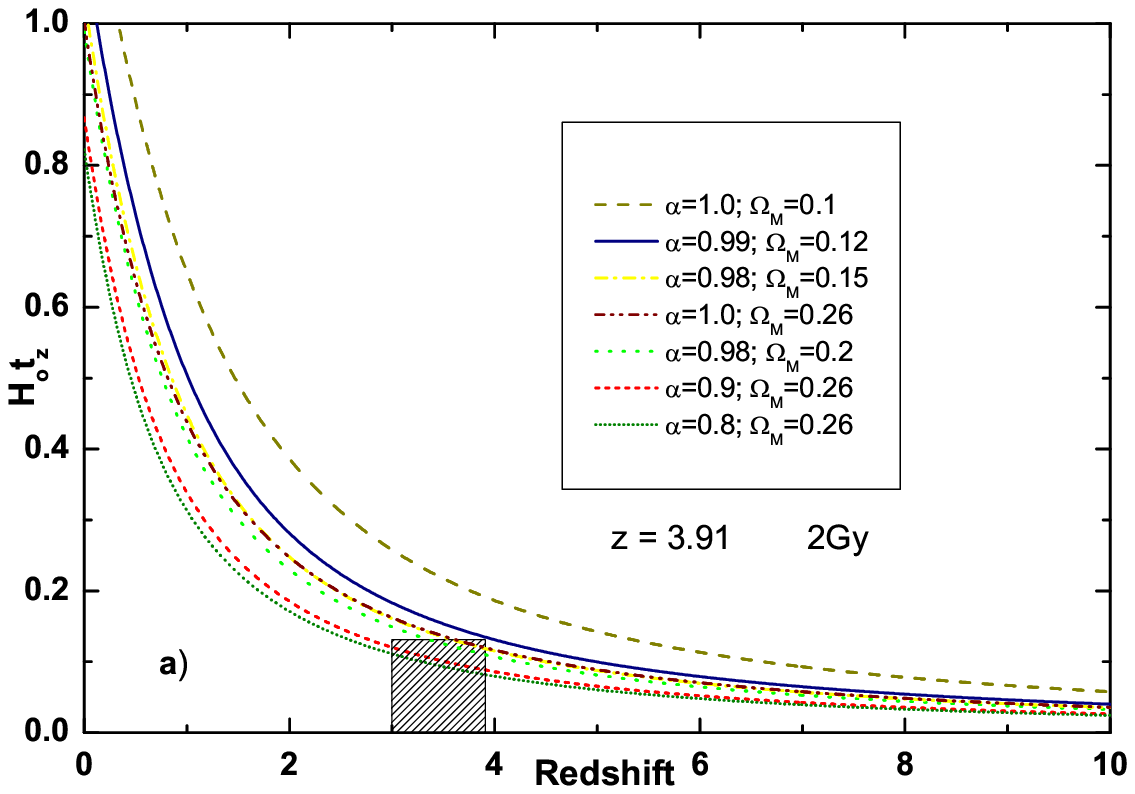}
\includegraphics[width=82mm]{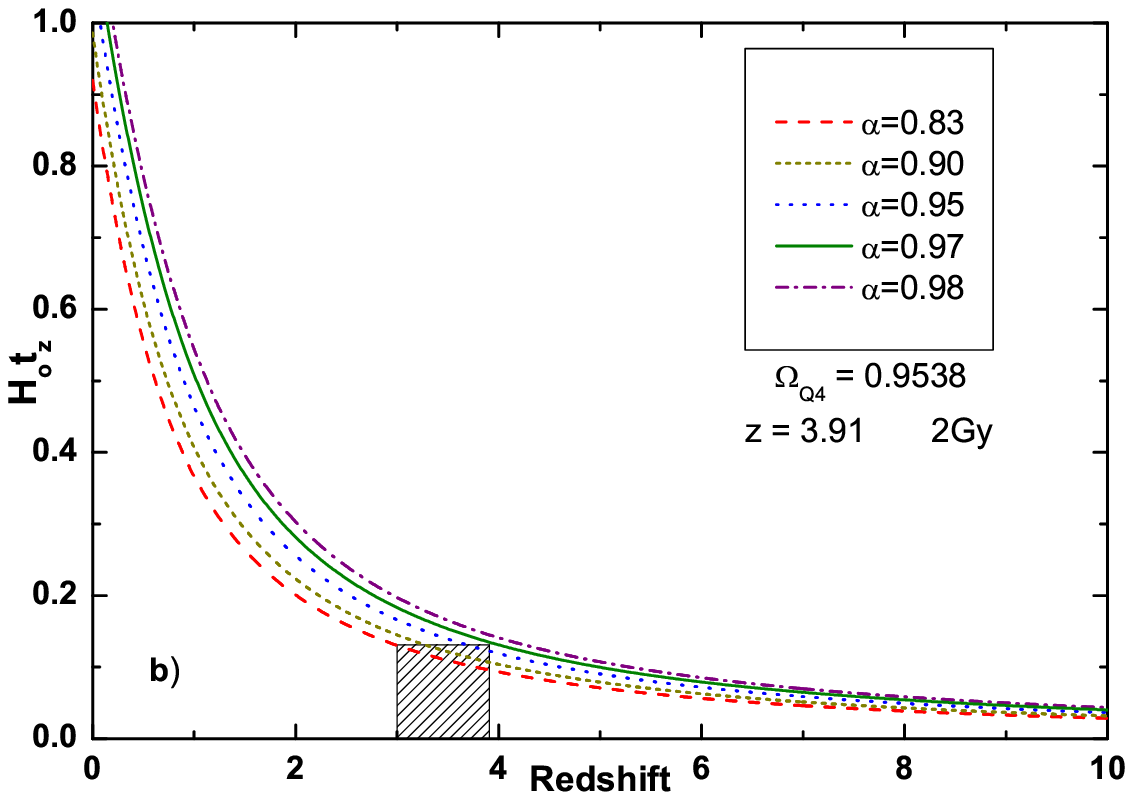}
\caption{High-$z$ Age Test. Panel {\bf a)} Dimensionless age parameter as a function of
redshift for some values of the pair ($\alpha,\Omega_{\rm M}$). Panel {\bf b)} Dimensionless age parameter as a function of redshift for some values of ($\alpha,\Omega_{\rm Q4}$). As explained
in the text, all curves crossing the shadowed area yield an age
parameter smaller than the minimal value required by the quasar APM
08279+5255, 2 Gyr, as reported by Hasinger et al. (2002) \cite{HasingerEtAl02} (see main text).\label{FigTz}}
\end{figure}


\begin{figure}[t]
\centering
\vspace{.2in}
\includegraphics[width=82mm]{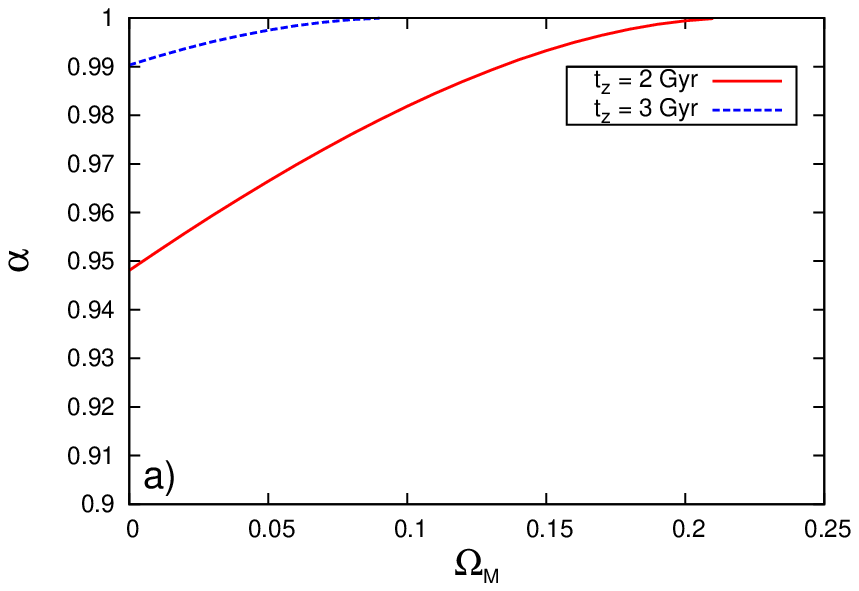}
\includegraphics[width=82mm]{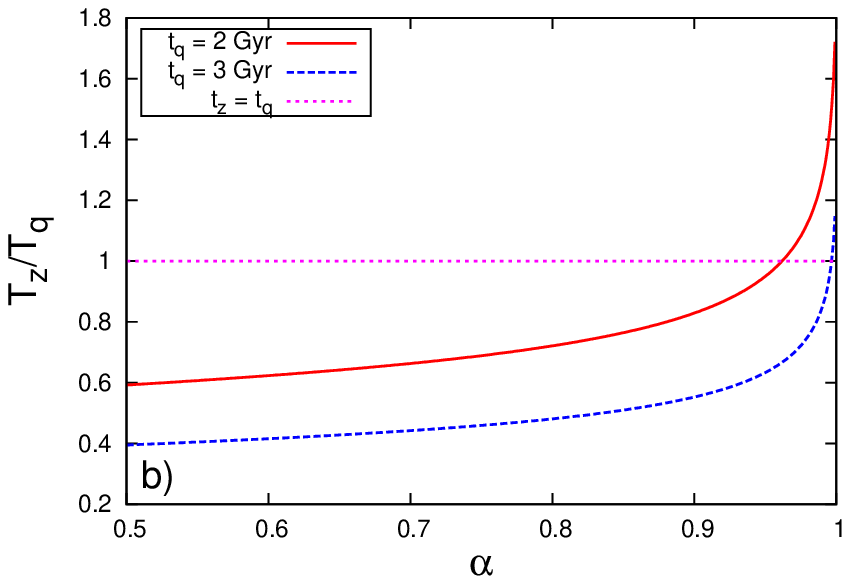}
\caption{Panel {\bf a)} Isochrones in the plane $\Omega_M$ {\it versus} $\alpha$, for the SC-gas Quintessence model, which match the estimated range of ages of the quasar. Panel {\bf b)} Ratio $T_z/T_q$ at the redshift of the quasar as a function of the $\alpha$ parameter showing the range of this parameter allowed by the existence of the quasar. For $t_q = 2$ Gyr, we have $\alpha > 0.965$ and for $t_q = 3$ Gyr, $\alpha > 0.997$.}
\label{FigIsoChro}
\end{figure}

In Fig. 1, we show the dimensionless age parameter $T_z = H_0 t_z$
as a function of the redshift for different values of
($\alpha,\Omega_{\rm M}$) in the Quintessence case, and, for some selected values of $\alpha$ in the Quartessence version.  The shadowed regions in the graphs were
determined from the minimal value of $T_q$. It means that any curve
crossing the rectangles yields an age parameter smaller than the
minimal value required by the presence of the quasar APM 08279+5255.
At this point, in line with the
arguments presented by Hasinger et al. (2002) \cite{HasingerEtAl02}, we recall that that X-ray
observations show a Fe/O ratio for the quasar APM 08279+5255 compatible with an age of 2 Gyr. Naturally, we do not expect such results to be free of
observational and/or theoretical uncertainties, but the analysis of the quoted reference has independently been confirmed by Fria\c{c}a and collaborators \cite{LA}.

In Fig. 2, we can see more clearly how the age of the quasar helps to constrain these models. In Panel {\bf (a)}, we show isochrones in the plane $\Omega_M-\alpha$ , for the Quintessence scenario.  Using the most conservative age estimate of the quasar (2 Gyr), the limits are $\Omega_M<0.21$  and $\alpha>0.947$. From Panel {\bf (b)} we  also see that the Quartessence scenario  is compatible with the presence of the quasar only if $\alpha>0.965$.  Therefore, as happens with other dark energy candidates \cite{LA}, the existence of APM 08279+5255 quasar constrains severely both formulations (Quintessence and Quartessence) of the Simplified Chaplygin Gas cosmology.    

\subsection{Age-Redshift and BAO: Joint Analysis}

\begin{figure}[t]
\centering
\includegraphics[width=82mm]{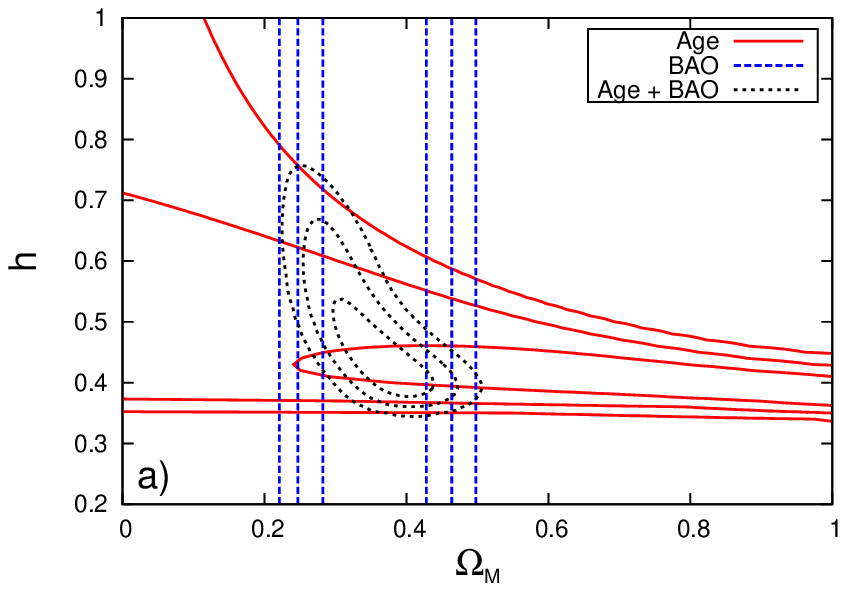}
\includegraphics[width=82mm]{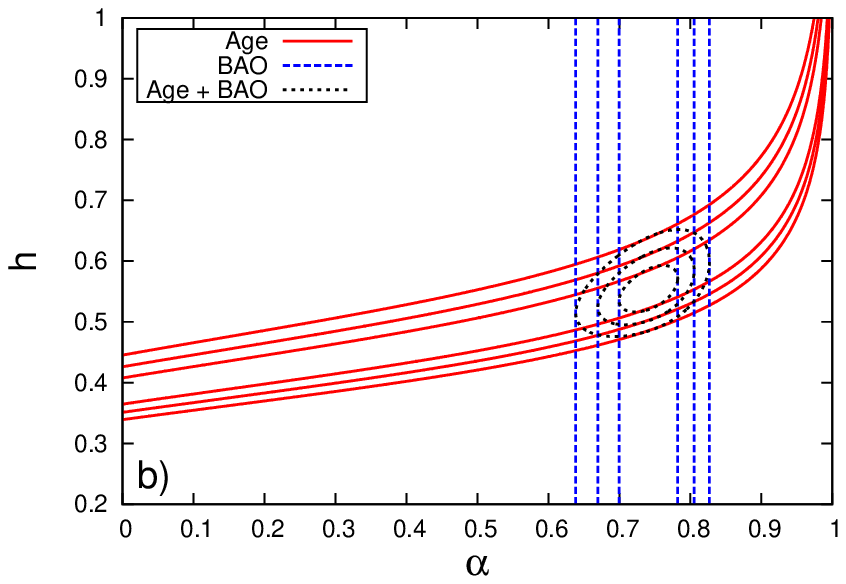}
\caption{Contours in the parameter spaces using the Age Redshift and BAO joint analysis. The contours correspond to 68.3\%, 95.4\% and 99.7\% confidence levels. Panel {\bf a)} Quintessence simplified model. The best fit parameters are $h=0.42^{+0.25}_{-0.060}$ and $\Omega_M=0.38^{+0.092}_{-0.13}$, at 95.4\% c.l. Panel {\bf b)} Quartessence simplified model. The best fits parameters are  $\alpha=0.743^{+0.073}_{-0.062}$ and $h=0.552^{+0.070}_{-0.057}$, at 95.4\% c.l. }
\label{FigCont}
\end{figure}

In order to break possible degeneracies in the simplified models, let us  now consider 
a joint analysis involving  a sample of 13 old galaxies \cite{McCa04,LimJesCun07} and data from the large scale structure (LSS). For the LSS data, we consider the recent measurements of the BAO peak in the large scale correlation function as inferred by Eisenstein {\it et al.} \cite{Eisenst05} using a large sample of luminous red galaxies from the SDSS Main Sample. The SDSS BAO measurement provides ${\cal A} =
0.469(n_s/0.98)^{0.35}\pm 0.017$. We shall fix the scalar spectral index as given by the the WMAP-5yr collaboration \cite{Komatsu08}, who find $n_s=0.960^{+0.014}_{-0.013}$. For constraining the parameters with basis on the age-redshift test we consider the oldest galaxies  in the samples observed by quoted  authors \cite{McCa04} and previously employed by Lima, Jesus and Cunha \cite{LimJesCun07} in the context of the cosmic concordance $\Lambda$CDM models. Actually, since the galaxy formation is a random process it is not need to consider 
the ages of all objects in order to constrain the model parameters. In other words, only the oldest ones are important for constraining a given cosmological model. It is also necessary  to take into account the incubation time ($t_{inc}$), that is, the time elapsed from the beginning of structure formation until the formation of the galaxy in question. It is reasonable to assume that such a quantity does not vary too much and that its uncertainty is quantified by $\sigma_{t_{inc}}$. In agreement with other works in the literature \cite{Fow86,Sand93}, we estimate that $t_{inc}=0.8\pm0.4$Gyr. 

Let us now perform a $\chi^2$ statistical analysis in order to constrain the free parameters of the models. In the Quintessence scenario me need to minimize the expression:
\begin{eqnarray}
\chi^2(\Omega_M,\alpha,h)=\chi^2_{Age}(\Omega_M,\alpha,h) + \chi^2_{BAO}(\Omega_M,\alpha) =\nonumber\\ =\sum_{i=1}^{13}\frac{{(t_{obs, i} + t_{inc} -
t_{th})^2}}{\sigma^2_i + \sigma^2_{inc}} + \left[\frac{{\cal A} -
0.469(n_s/0.98)^{0.35}}{0.017}\right]^2,
\end{eqnarray}
while in the Quartessence  we have to minimize:
\begin{eqnarray}
\chi^2(\alpha,h)=\chi^2_{Age}(\alpha,h) + \chi^2_{BAO}(\alpha) 
=\sum_{i=1}^{13}\frac{{(t_{obs, i} + t_{inc} -
t_{th})^2}}{\sigma^2_i + \sigma^2_{inc}} + \left[\frac{{\cal A} -
0.469(n_s/0.98)^{0.35}}{0.017}\right]^2,
\end{eqnarray}
since the baryon contribution can be fixed by the WMAP-5yr ($\Omega_b=0.0462\pm0.0015$), and, as such, only the pair of parameters  ($\alpha, h$) can be considered in the Quartessence scenario. 

To begin with let us consider the Quintessence case. Constraints on the free parameters are obtained by marginalizing the associated likelihood expression over $\alpha$ as follows:
\begin{equation}
 \tilde{\cal L}(\Omega_M,h)=\int_{-\infty}^{+\infty}\pi(\alpha){\cal L}(\Omega_M,\alpha,h)d\alpha=\int_0^1{\cal L}(\Omega_M,\alpha,h)d\alpha,
\end{equation}
where ${\cal L}$ is the likelihood, given by ${\cal L}\propto e^{-\chi^2/2}$, and $\pi(\alpha)$ is the prior on $\alpha$, assumed to be a top-hat on the physical region $0<\alpha\le1$.  In Panel {\bf (a)} of Fig. \ref{FigCont} we display  the constraints on the resulting 2-dimensional plane $\Omega_M-h$. Naturally, no marginalization is necessary to the  Quartessence version since it has only two free parameters. In this case, the required constraints in the  plane $\alpha-h$ are displayed in Panel {\bf (b)} of Fig. \ref{FigCont}.

In Tab. \ref{Tab1} we display some estimates of $H_0$. In the standard model of a flat
$\Lambda$-dominated universe with CMB data alone, it is found $H_0=71.9^{+2.6}_{-2.7}$.
On Sandage et al. (2006) \cite{Sand06}, the final result of the HST collaboration, ranging over 15 years, it is found $H_0$(cosmic)$ = 62.3 \pm 1.3$ (random) $\pm 5.0 $(systematic),
based on 62 SNe Ia with $3000 < v_{CMB} < 20 000 {\rm kms^{-1}}$ and on 10
luminosity-calibrated SNe Ia. Their local value of $H_0$ ($300 < v_{220}
< 2000 {\rm kms^{-1}}$) is $H_0$(local) $= 60.9\pm 1.3 $(random) $\pm 5.0$
(systematic), from 25 Cepheid and 16 SNe Ia distances, involving a
total of 34 different galaxies. This result is in disagreement with the result of Freedman et al. (2001) \cite{Freedman01}, $H_0=72\pm8$, obtained from the HST Key Project, and this discrepancy is a matter of debate. In what concerns our analysis here, our principal results are, for the S-CG, $0.26 \leq \Omega_M \leq 0.47$, $0.36 \leq h \leq 0.67$
($2\sigma$) and for the Q4-CG, $0.681 \leq \alpha \leq 0.805$, $0.495 \leq h \leq
0.622$ ($2\sigma$). These results depend weakly on the incubation
time ($t_{inc} = 0.8 \pm 0.4$ Gyr). On Tab. \ref{Tab1}, we may see that our result on the Hubble constant is consistent with the value of Sandage et al. (2006) \cite{Sand06}, where was
found $H_0 = 62.3 \pm 1.3$ km s$^{-1}$ Mpc$^{-1}$ (statistical).

\begin{table}[t]
{\begin{tabular}{c c c} \toprule
\multicolumn{1}{c}{Method}& \multicolumn{1}{c}{Reference}&
\multicolumn{1}{c}{$H_0 $}\\ \colrule
 Age Redshift & Jimenez et al. (2003) \cite{Jime03}& $ 69 \pm 1.2$
\\S-Z effect & Schmidt et al. (2004) \cite{Schi04}& $ 69^{+8}_{-8}$
\\ S-Z effect & Jones et al. (2005) \cite{Jon05} & $ 66^{+11 +9}_{-10 -8}$
\\ SNe Ia/Cepheid & Freedman et al. (2001) \cite{Freedman01} & $ 72 \pm 8$
\\ SNe Ia/Cepheid & Sandage et al. (2006) \cite{Sand06} & $ 62 \pm 1.3\pm 5$
\\ CMB & Dunkley et al. (2008) \cite{Komatsu08} & $ 71.9^{+2.6}_{-2.7}$
\\ Old Galaxies+BAO & Lima et al. (2007) \cite{LimJesCun07} & $71 \pm 4$
\\ Old Galaxies+BAO & Figure \ref{FigCont}. {\bf a)} (S-CG) & $ 42^{+25}_{-6}$ (95\% c.l.) 
\\ Old Galaxies+BAO & Figure \ref{FigCont}. {\bf b)} (Q4-CG)&  $ 55.2^{+7.0}_{-5.7}$ (95\% c.l.) 
\\ \botrule
\end{tabular}}
\caption{Limits on $H_0$ ${\rm (km s^{-1} Mpc^{-1})}$ using different methods.}
\label{Tab1}
\end{table}

\section{Results and discussions}

It is usually believed that old high-redshift objects may play an important
role to the question related to the ultimate fate of the Universe.
In a point of fact, their age estimates provide a powerful technique for constraining
the free parameters in a given cosmological model \cite{Krauss97,LA}. In particular, this means that the so-called
high-redshift  age crisis is now becoming an important complement
to other independent cosmological tests. 

In this work we have discussed some constraints in the so-called Simplified Chaplygin Gas (SC-gas) both in the Quintessence and Quartessence versions. 
Initially, we have investigated the dimensionless age parameter $T_z = H_0 t_z$. In the simplified Quintessence  scenario, by assuming that the quasar has an age of at least a 2 Gyr,  we have obtained  $\alpha\geq 0.947$ while for the simplified Quartessence model the limit is $\alpha>0.965$. We have also studied the constraints on the
Hubble constant using a joint analysis involving  different OHRGs
samples and the current SDSS measurements of the baryon acoustic
peak for both scenarios. In this case, we have found  values of the Hubble constant moderately low, specially, to the  Quintessence case (see Fig. 3 and Table 1).

As a conclusion, we would like to stress that our results point to at least 3 possibilities. The first one is that the simplified Chaplygin type gas models (Quintessence and Quartessence) do not provide a good fit to the Age+BAO data because the best fit predicted for $H_0$ is low. A second possibility is that the estimated ages of the old high redshift galaxies considered here (or even their incubation time) need to be revised, and this may increase the predicted value of the Hubble constant. Finally, as has been advocated by some authors \cite{Sand06,Blanch}, there exist the possibility that we live in a Universe with Hubble constant smaller than the one predicted by the HST Key Project \cite{Freedman01}.

\section*{Acknowledgments}
We are grateful to J. V. Cunha, J. A. S. Lima and S. H. Pereira for helpful discussions. 
RCS and JFJ are supported by CNPq (Brazilian Research Agency).


\begin{thebibliography}{00}   

\bibitem{Riess} P. Astier {\it et al.}, Astron. Astrophys. {\bf 447}, 31 (2006); A. G. Riess {\it et al.}, Astrophys. J. {\bf{659}}, 98 (2007); T. M. Davis {\it et al.}, Astrophys. J. {\bf{666}}, 716 (2007); M. Kowalski {\it et al.}, arXiv:0804.4142 [astro-ph]. 


\bibitem{Sperg07} D. N. Spergel {\it et al.}, Suppl. {\bf{148}}, 175 (2003); D. N. Spergel {\it et al.}, Astrophys. J.  Supl. {\bf 170}, 377 (2007).

\bibitem{OzTa87}  M. $\ddot{\rm{O}}$zer and M. O. Taha, Phys. Lett. B {\bf 171}, 363 (1986);  J. A. S. Lima and J. M. F. Maia,   Phys. Rev. {\bf D49}, 5597 (1994);  J. A. S. Lima and J. C. Carvalho, Gen. Rel. Grav. {\bf 26}, 909 (1994); J. A. S. Lima, Phys. Rev. {\bf D54}, 2571 (1996), [gr-qc/9605055]; J. M. Overduin and F. I. Cooperstock, Phys. Rev. {\bf D58}, 043506 (1998);  J. S. Alcaniz and J. A. S. Lima, Phys. Rev. {\bf D72}, 063516 (2005), [astro-ph/0507372]; J. F. Jesus {\it et al.}, arXiv:0806.1366 [astro-ph].

\bibitem{turner97} M. S. Turner and M. White, Phys. Rev. {\bf D56}, R4439 (1997); T. Chiba, N. Sugiyama and  T. Nakamura, Mon. Not. R. Astron. Soc.
{\bf 289}, L5 (1997); J. S. Alcaniz and  J. A. S. Lima,   Astrophys.
J.  {\bf 521}, L87 (1999),[astro-ph/9902298]; {\bf ibdem} Astrophys.  J.  {\bf 550},
L133 (2001), [astro-ph/0101544];  J. A. S. Lima, J. V. Cunha and J. S. Alcaniz, Phys.
Rev. D {\bf 68}, 023510 (2003), [astro-ph/0303388]; M. P. D\c{a}browski, [arXiv:gr-qc/0701057] (2007).
\bibitem{PR03}  R. R. Caldwell and P. J. Steinhardt,  Phys. Rev. D {\bf
57}, 6057 (1998); A. Alam, V. Sahni and  A. A. Starobinsky, JCAP
{\bf 0406}, 008 (2004); F. C. Carvalho {\it et al.}, Phys. Rev. Lett. {\bf 97}, 081301 (2006), [astro-ph/0608439]; {\bf ibdem}, [arXiv:0704.3043] (2007); J. V. Cunha, L. Marassi and R. C. Santos, IJMPD {\bf 16}, 403 (2007), S. Nesseris and L. Perivolaropoulos, JCAP 0701, 018 (2007); R. C. Santos and J. A. S. Lima, Phys. Rev. {\bf D77}, 023519 (2008), arXiv:0803.1865 [astro-ph]; J. A. S. Lima and S. H. Pereira, arXiv:0801.0323 [astro-ph]. 
  

\bibitem{kamen} A. Kamenshchik, U. Moschella and V. Pasquier, Phys. Lett. B {\bf 511}, 265 (2001); M. C. Bento, O. Bertolami and A. Sen, Phys. Rev. D {\bf 66}, 043507 (2002); N. Bil\'{\i}c, G. B. Tupper and R. D. Viollier, Phys. Lett. B {\bf 535}, 17 (2002); J. V. Cunha, J. S. Alcaniz and J. A. S. Lima, Phys. Rev. D {\bf 69}, 083501 (2004), [astro-ph/0306319];  J. S. Alcaniz and J. A. S. Lima, Astrophys. J. {\bf 618}, 16 (2005), [astro-ph/0308465]; P. Wu and H. Yu Class. Quant. Grav. {\bf 24}, 4661 (2007).

\bibitem{review} T. Padmanabhan, Phys. Rept. {\bf 380}, 235 (2003); P. J. E. Peebles and  B. Ratra,  Rev. Mod. Phys. {\bf 75}, 559 (2003);
J. A. S. Lima, Braz. Jour. Phys. {\bf 34}, 194 (2004),
[astro-ph/0402109]; E. J. Copeland, M. Sami and S. Tsujikawa, Int. J. Mod. Phys. {\bf{D15}}, 1753 (2006); M. S. Turner and D. Huterer, [arXiv:0706.2186] (2007).

\bibitem{old} J. S. Dunlop {\it et al.}, {\it Nature} {\bf 381}, 581 (1996); H. Spinrad {\it et al.}, {\it ApJ} {\bf 484}, 581 (1997). 

\bibitem{Krauss97} L. Krauss, {\it ApJ} {\bf 480}, 466 (1997);
J. S. Alcaniz and J. A. S. Lima, {\it ApJ }{\bf 521}, L87 (1999); {\bf ibdem}, ApJ {\bf 550}, L133 (2001); J. A. S. Lima and J. S. Alcaniz, {\it  MNRAS} {\bf 317}, 893 (2000), [astro-ph/0005441].


\bibitem{HasingerEtAl02} G. Hasinger, N. Schartel and S. Komossa, Astrophys. J. {\bf 573}, L77 (2002).

\bibitem{LA} J. S. Alcaniz, J. A. S. Lima, and J. V. Cunha, {\it
MNRAS} {\bf 340}, L39 (2003), [astro-ph/0301226]; J. V. Cunha and R. C. Santos, Int. J. Mod. Phys. {\bf D13}, 1321 (2004), [astro-ph/0402169]; A. C. S. Fria\c{c}a, J. S. Alcaniz and J. A. S. Lima, MNRAS {\bf 362}, 1295 (2005), [astro-ph/0504031];  D. Jain and  A. Dev, Phys. Lett. B {\bf 633}, 436 (2006); J. F. Jesus, [astro-ph/0603142].   

\bibitem{Age} L. A. Nolan {\it et al.}, MNRAS {\bf 323}, 308 (2001); R. Jimenez and A. Loeb, ApJ {\bf 573}, 37 (2002); R. Jimenez {\it et al.}, ApJ {\bf 593}, 622 (2003); S. Weinberg, Cosmology, Oxford UP, New York (2008).

\bibitem{Eisenst05} D. J. Eisenstein {\it et al.}, Astrophys. J. {\bf 633}, 560 (2005).

\bibitem{CLM07} J. V. Cunha, L. Marassi and J. A. S. Lima, Mon. Not. R. Ast. Soc., {\bf 379}, L1 (2007), [astro-ph/0611934]. 

\bibitem{scg} J. A. S. Lima, J. V. Cunha, J. S. Alcaniz, astro-ph/0608469 (2006).

\bibitem{scg1} J. A. S. Lima, J. V. Cunha, J. S. Alcaniz, astro-ph/0611007 (2006). 

\bibitem{GCG} P. P. Avelino, L. M. G. Be\c{c}a, J. P. M. de Carvalho, C.
J. A. P. Martins and P. Pinto, Phys. Rev. {\bf D67}, 023511
(2003); M. Makler, S. Q. de Oliveira and I. Waga, Phys. Lett. {\bf B
68}, 123521 (2003); A. Dev, J. S. Alcaniz and D. Jain, Phys. Rev {\bf D67},
023515 (2003), astro-ph/0209379; P. Wu and  H. W. Yu, Phys. Lett. {\bf B644} 16 (2007); R. A. Sussman, arXiv:0801.3324 [gr-qc]. 

\bibitem{Peebles} P. J. E. Peebles, Principles of Physical Cosmology, Princeton UP, New Jersey (1993).  

\bibitem{Freedman01} W. L. Freedman {\it et al.}, Astrophys. J. {\bf 553}, 47 (2001).

\bibitem{McCa04} P. J. McCarthy  {\it et al.}, ApJ Letters {\bf 614}, 9 (2004); N. D. Roche {\it et al.}, MNRAS {\bf 370}, 74 (2006); Longhetti, M. et al., MNRAS {\bf 374}, 614 (2007).

\bibitem{LimJesCun07} J. A. S. Lima, J. F. Jesus and J. V. Cunha, arXiv:0709.2195 [astro-ph].

\bibitem{Komatsu08} E. Komatsu et al., arXiv:0803.0547 (2008); J. Dunkley et al., arXiv:0803.0586 (2008).

\bibitem{Fow86} Fowler, W. A., Q. Jl. R. Astr. Soc. {\bf 28}, 87 (1987).

\bibitem{Sand93} Sandage, A., Astron. J. {\bf 106}, 719 (1993).

\bibitem{Jime03} R. Jimenez, {\it et al.} Astrophys. J. {\bf 593}, 622 (2003). 

\bibitem{Kunz04} M. Kunz, P. S. Corasaniti, D. Parkinson, E. J. Copeland, Phys.
Rev. D {\bf 70}, 041301 (2004).

\bibitem{Schi04} R. W. Schmidt, S. W. Allen and A. C. Fabian, Mon. Not. R. Astron. Soc. {\bf 352}, 1413 (2004).

\bibitem{Jon05} Jones, M. E. {\it et al.}, Mon. Not. R. Astron. Soc. {\bf 357}, 518 (2005).

\bibitem{Sand06} A. Sandage  {\it et al.}, Astrophys. J. {\bf  653}, 843-860 (2006).
 
\bibitem{Blanch} A. Blanchard, M. Douspis, M. Rowan-Robinson, and S. Sarkhar, Astron. and Astrophys. {\bf 412}, 35 (2003); 
S. Sarkhar (2007), [astro-ph0710.5307v1]. 

\end{thebibliography}
\end{document}